\documentstyle[12pt]{article}

\begin{document} 
\vskip 3cm
\begin{center}
{\large Normal-state pseudogap and electron flat dispersion in copper
oxide materials} \\
\medskip
Feng Yuan \\
Department of Physics, Beijing Normal University, Beijing
100875, China \\
Shiping Feng \\
CCAST (World Laboratory) P. O. Box 8730, Beijing 100080, China and\\
$^{*}$Department of Physics, Beijing Normal University, Beijing
100875, China and \\
National Laboratory of Superconductivity, Academia Sinica, Beijing
100080, China \\
\end{center}
\bigskip

The anomalous momentum and doping dependence of the electron spectral
function and electron dispersion for copper oxide materials in the
underdoped regime are studied within the $t$-$J$ model. It is shown
that the electron spectrum is changed with dopings, and the electron
dispersion exhibits a flat band around ($\pi$,0) point in the
Brillouin zone, which leads to the normal-state pseudogap formation.
The theoretical results are consistent with the experiments and
numerical simulations.\\
\leftline{71.27.+a, 73.20.Dx, 74.72.-h,74.25.Jb}
\newpage

The copper oxide materials are unusual in that the undoped materials
are antiferromagnetic (AF) insulators, and changing the carrier
concentration by ionic substitution or increase of the oxygen content
turns these compounds into correlated metals leaving short-range AF
correlations still intact \cite{n1,n2}, where a central issue to
clarify the nature of the anomalous properties is how the electronic
structure evolves with hole dopings, since many of the unusual
physical properties including the anomalously high superconducting
transition temperature have often been attributed to particular
characteristics of low energy excitations determined by the electronic
structure \cite{n1,n2,n3,n4}. The experimental measurements from the
angle-resolved photoemission spectroscopy \cite{n5,n6,n7,n8} show that
electron spectrum $A({\bf k},\omega)$ in copper oxide materials is
strongly momentum and doping dependent, and has an anomalous form as
a function of energy $\omega$ for ${\bf k}$ in the vicinity of
(${\pi}$,0) point in the Brillouin zone, which leadd to the flat band
near momentum ($\pi$,0) with anomalously small changes of electron
energy as a function of momentum. This flat band around ($\pi$,0)
point reflects the underlying electronic structure near a band saddle
point, and has a particular importance in the mechanism of the
normal-state pseudogap formation, since in the underdoped regime the
normal-state pseudogap starts growing first in the single-particle
excitations around ($\pi$,0) point, and then exists in a wide range
of dopings \cite{n6,n9,n10,n11}. This flatted band also makes
degenerate excitations and may cause various instabilities. It is
believed that the broad feature in the spectrum around ($\pi$,0) point
in copper oxide materials in the underdoped regime is manifestation of
a strong coupling between the quasiparticle and collective excitations
\cite{n12}. On the other hand, it has been argued that there is the
intriguing connection between the velocity scale $v^{*}\ll v_{F}$ and
the flat band \cite{n13}, where $v_{F}$ is the Fermi velocity, while
$v^{*}$ has been found from the inelastic neutron-scattering
experiments that it is related to the superconducting transition
temperature $T_{c}$ by the simple relation as
$K_{B}T_{c}=hv^{*}\lambda$, where $\lambda$ is the splitting of the
incommensurate peaks or the peak width. Although this intriguing
connection is not unambiguously confirmed by experiments, we still
believe that the flat band may be an essential element to understanding
of the superconducting mechanism in copper oxide materials.

The electron spectrum and flat band in copper oxide materials have
been extensively studied theoretically within some strongly
correlated models \cite{n14,n15,n16,n17,n18}. The most striking
aspect is the presence of the flat band that can not be explained by
either of the band theory scenarios. The numerical calculation of the
electron spectrum based on the two-dimensional (2D) large U Hubbard
\cite{n14,n15} model shows the flat band similar to those observed
in experiments. In these calculation, the flat band arises from the
large Coulomb interaction U. The flat band has been also observed
in the 2D $t$-$J$ model \cite{n16,n17,n18}, where the quantitative
agreement between the experiment and numerical simulation along the
(0,0) to ($\pi,\pi$) direction is significant because it shows
unambiguously that the energy scale of the insulating band is
controlled by the magnetic interaction $J$. Moreover, many authors
\cite{n19} suggest that the flat band is a consequence of spin and
charge excitations. Although the exact origin of the flat band saddle
point still is controversial, a strongly correlated many-body like
approach may be appropriate to describe the electronic structure of
copper oxide materials. To shed light on this issue, we, in this
paper, try to study the momentum and doping dependence of the
electron spectrum within the framework of the fermion-spin theory
\cite{n20}. Our results show that the electron spectrum is changed
with dopings, and the electron dispersion exhibits the flat band
around ($\pi$,0) point, which leads to the normal-state pseudogap
formation.

Very soon after the discovery of copper oxide superconductors, many
authors \cite{n21} suggested that the essential physics of these
materials is contained in doped Mott insulators, which may be
effectivelly described by the 2D $t$-$J$ model acting on the space
with no doubly occupied sites. On the other hand, there is a lot of
evidence from the experiments and numerical simulations in favour of
the $t$-$J$ model as the basic underlying microscopic model \cite{n22}.
The $t$-$J$ model on a square lattice can be written as,
\begin{eqnarray}
H = -t\sum_{\langle ij\rangle\sigma}C^{\dagger}_{i\sigma}C_{j\sigma}
+ h.c. - \mu \sum_{i\sigma}C^{\dagger}_{i\sigma}C_{i\sigma} +
J\sum_{\langle ij\rangle}{\bf S}_{i}\cdot {\bf S}_{j} ,
\end{eqnarray}
supplemented by the on-site local constraint $\sum_{\sigma}
C^{\dagger}_{i\sigma}C_{i\sigma}\leq 1$ to avoid the double
occupancy, where $C^{\dagger}_{i\sigma}$ ($C_{i\sigma}$) are the
electron creation (annihilation) operators, ${\bf S}_{i}=
C^{\dagger}_{i}{\bf \sigma} C_{i}/ 2$ are spin operators with
${\bf \sigma}=(\sigma_{x},\sigma_{y},\sigma_{z})$ as Pauli matrices,
$\mu$ is the chemical potential, and the summation $\langle ij\rangle$
is carried over nearest nonrepeated bonds. Since the $t$-$J$ model was
originally introduced as an effective Hamiltonian of the large-U
Hubbard model \cite{n21}, where the on-site Coulomb repulsion U is
very large as compared with the electron hopping energy $t$, which
leads to that electrons become strongly correlated to avoid double
occupancy, therefore the strong electron correlation in the $t$-$J$
model manifests itself by the electron single occupancy on-site local
constraint. This is why the crucial requirement is to impose this
electron on-site local constraint for a proper understanding of the
physics of copper oxide materials \cite{n23}. To incorporate this
local constraint, the fermion-spin theory based on the charge-spin
separation has been proposed \cite{n20}. In the fermion-spin theory,
the constrained electron operators in the $t$-$J$ model are
decomposed as,
\begin{eqnarray}
C_{i\uparrow}=h^{\dagger}_{i}S^{-}_{i},~~~~
C_{i\downarrow}=h^{\dagger}_{i}S^{+}_{i},
\end{eqnarray}
with the spinless fermion operator $h_{i}$ keeps track of the charge
(holon), while the pseudospin operator $S_{i}$ keeps track of the
spin (spinon). The main advantage of this approach is that the
electron on-site local constraint can be treated exactly in
analytical calculations. In this case, the low-energy behavior of
the $t$-$J$ model (1) in the fermion-spin representation can be
written as,
\begin{eqnarray}
H = t\sum_{i\hat{\eta}}h^{\dagger}_{i+\hat{\eta}}h_{i}
(S^{+}_{i}S^{-}_{i+\hat{\eta}}+S^{-}_{i}S^{+}_{i+\hat{\eta}})
+ \mu \sum_{i}h^{\dagger}_{i}h_{i} + J_{eff}\sum_{i\hat{\eta}}
({\bf S}_{i}\cdot {\bf S}_{i+\hat{\eta}}),
\end{eqnarray}
where $\hat{\eta}=\pm\hat{x}$, $\pm\hat{y}$, $J_{eff}=
J[(1-\delta)^{2}-\phi ^{2}]$, the holon particle-hole parameter
$\phi=\langle h^{\dagger}_{i}h_{i+\hat{\eta}}\rangle$, and
$S^{+}_{i}$ and $S^{-}_{i}$ are the pseudospin raising and
lowering operators, respectively. As a consequence, the kinetic
part in the $t$-$J$ model has been expressed as the holon-spinon
interaction in the fermion-spin representation, which dominates the
physics in the underdoped and optimally doped regimes in copper oxide
materials \cite{n24,n25}. In this paper, we hope to discuss the
electronic structure of copper oxide materials, and therefore it
needs to calculate the electron Green's function $G(i-j,t-t')=\langle
\langle C_{i\sigma}(t);C^{\dagger}_{j\sigma}(t')\rangle\rangle$.
According the fermion-spin transformation (2), the electron Green's
function is a convolution of the spinon Green's function $D(i-j,t-t')
=\langle\langle S^{+}_{i}(t);S^{-}_{j}(t')\rangle\rangle$ and holon
Green's function $g(i-j,t-t')=\langle\langle h_{i}(t)h^{\dagger}_{j}
(t')\rangle\rangle$, and can be formally expressed in terms of the
spectral representation as,
\begin{eqnarray}
G({\bf k},\omega)={1\over N}\sum_{q}\int^{\infty}_{-\infty}
{d\omega' \over 2\pi}\int^{\infty}_{-\infty}{d\omega'' \over 2\pi}
A_{h}({\bf q}, \omega')A_{s}({\bf q+k},\omega''){n_{F}(\omega')+
n_{B}(\omega'')\over \omega+\omega'-\omega''},
\end{eqnarray}
where the holon spectral function $A_{h}({\bf q},\omega)=
-2{\rm Im}g({\bf q},\omega)$, the spinon spectral function
$A_{s}({\bf k},\omega)=-2{\rm Im}D({\bf k},\omega)$, and
$n_{B}(\omega)$ and $n_{F}(\omega)$ are the boson and fermion
distribution functions for spinons and holons, respectively. In
this case, the electron spectral function $A({\bf k},\omega)
=-2{\rm Im}G({\bf k},\omega)$ can be obtained as,
\begin{eqnarray}
A({\bf k},\omega)={1\over N}\sum_{q}\int^{\infty}_{-\infty}
{d\omega' \over 2\pi}A_{h}({\bf q},\omega')A_{s}({\bf q+k},
\omega'+\omega)[n_{F}(\omega')+n_{B}(\omega'+\omega)].
\end{eqnarray}

Within the fermion-spin theory, the mean-field theory in the
underdoped and optimally doped regimes without AF long-range-order
(AFLRO) has been developed \cite{n26}, and the mean-field spinon
and holon Green's functions $D^{(0)}({\bf k}, \omega)$ and $g^{(0)}
({\bf k},\omega)$ have been evaluated as,
\begin{eqnarray}
D^{(0)}({\bf k},\omega)={B_{k}\over 2\omega_{k}}\left ({1\over
\omega -\omega_{k}}-{1\over \omega +\omega_{k}}\right ),
\end{eqnarray}
\begin{eqnarray}
g^{(0)}({\bf k},\omega)={1\over \omega-\xi_k},
\end{eqnarray}
respectively, where $B_{k}=\Lambda [(2\epsilon\chi_{z}+\chi)
\gamma_{k}-(\epsilon\chi+2\chi_{z})]$, $\gamma_{{\bf k}}=(1/Z)
\sum_{\eta}e^{i{\bf k}\cdot\hat{\eta}}$, $\Lambda =2ZJ_{eff}$,
$\epsilon=1+2t\phi/J_{eff}$, $Z$ is the number of the nearest
neighbor sites, while the mean-field spinon spectrum
\begin{eqnarray}
\omega^{2}_{k}=\Lambda^{2}\left (\alpha\epsilon[\chi_{z}\gamma_{k}
+{1\over 2Z}\chi ]-[\alpha C_{z}+{1\over 4Z}(1-\alpha)]\right )
(\epsilon\gamma_{k}-1) \nonumber \\
+\Lambda^{2}\left (\alpha\epsilon[{1\over 2}\chi\gamma_{k}+
{1\over Z}\chi_{z}]-{1\over 2}\epsilon[\alpha C+{1\over 2Z}
(1-\alpha)]\right )(\gamma_{k}-\epsilon),
\end{eqnarray}
and the mean-field holon spectrum $\xi_k=2Z\chi t\gamma_k+\mu$,
with the spinon correlation functions
$\chi=\langle S_{i}^{+}S_{i+\hat{\eta}}^{-}\rangle$,
$\chi_{z}=\langle S_{i}^{z}S_{i+\hat{\eta}}^{z}\rangle$,
$C=(1/Z^{2})\sum_{\hat{\eta},\hat{\eta'}}\langle
S_{i+\hat{\eta}}^{+}S_{i+\hat{\eta'}}^{-}\rangle$, and
$C_{z}=(1/Z^{2})\sum_{\hat{\eta}, \hat{\eta'}}\langle
S_{i+\hat{\eta}}^{z}S_{i+\hat{\eta'}}^{z}\rangle$. In order not to
violate the sum rule of the correlation function $\langle
S^{+}_{i}S^{-}_{i}\rangle=1/2$ in the case without AFLRO, the
important decoupling parameter $\alpha$ has been introduced in the
mean-field calculation, which can be regarded as the vertex correction
\cite{n26}. The mean-field order parameters $\chi$, $C$, $\chi_z$,
$C_z$, $\phi$ and chemical potential $\mu$ have been determined by the
self-consistent calculation. Based on this mean-field theory, the
electron mean-field spectral function and electron dispersion have
been discussed \cite{n26}, where the most important feature is that
the mean-field intensity peaks in the electron spectral function
are qualitatively consistent with the numerical simulations. However,
the detailed behaviors of the flat band near ($\pi$,0) point and
normal-state pseudogap related to this flat band should be studied
beyond the mean-field approximation, since these behaviors are
associated with fluctuations of holons and spinons (then electrons).

The spinon and holon may be separated at the mean-field level, but
they are strongly coupled beyond the mean-field approximation due
to many-body correlations. In this paper, we limit the spinon part
to the first-order (mean-field level) since in the underdoped and
optimally doped regimes without AFLRO, the spinon magnetic energy is
much smaller than the holon kinetic energy, and some physical
properties can be well described at this level. On the other hand, it
has been shown that there is a connection between the charge dynamics
(the anomalously temperature dependence of the resistivity) and the
saddle-point singularity around ($\pi$,0) point and normal-state
pseudogap  \cite{n4,n9,n27,n28}. We \cite{n24,n25} have discussed the
charge dynamics of copper oxide materials within the fermion-spin
theory, and found that there is no direct contribution to the charge
dynamics from spinons, although the strongly correlation between
holons and spinons has been considered through the spinon's order
parameters entering in the holon part of the contribution to the
charge dynamics. Therefore we treat the holon part by using the loop
expansion to the second-order correction as in the discussion of the
charge dynamics. In this case, the holon self-energy due to the
spinon pair bubble has been obtained \cite{n24} as,
\begin{eqnarray}
\Sigma_{h}^{(2)}({\bf k},i\omega_{n})=(Zt)^{2}{1\over N^2}
\sum_{pp'}(\gamma_{p'-k}+\gamma_{p'+p+k})^{2}{B_{p'}B_{p+p'}
\over 4\omega_{p'}\omega_{p+p'}}\times \nonumber \\
\left ( 2{n_{F}(\xi_{p+k})[n_{B}(\omega_{p'})-n_{B}(\omega_{p+p'})]
-n_{B}(\omega_{p+p'})n_{B}(-\omega_{p'})\over \omega+\omega_{p+p'}-
\omega_{p'}-\xi_{p+k}} \right. \nonumber \\
+{n_{F}(\xi_{p+k})[n_{B}(\omega_{p+p'})-n_{B}(-\omega_{p'})]+n_{B}
(\omega_{p'})n_{B}(\omega_{p+p'})\over \omega+\omega_{p'}+
\omega_{p+p'}-\xi_{p+k}} \nonumber \\
\left. -{n_{F}(\xi_{p+k})[n_{B}(\omega_{p+p'})-n_{B}(-\omega_{p'})]
-n_{B}(-\omega_{p'})n_{B}(-\omega_{p+p'})\over \omega-\omega_{p+p'}
-\omega_{p'}-\xi_{p+k}}\right ),
\end{eqnarray}
while the full holon propagator $g({\bf k},\omega_{n})$ can be
expressed in terms of the self-energy (9) as,
\begin{eqnarray}
g({\bf k},\omega)={1\over g^{(0)-1}({\bf k},\omega)-\Sigma_{h}^{(2)}
({\bf k},\omega)}={1\over \omega-\xi_{k}-\Sigma_{h}^{(2)}({\bf k}.
\omega)}.
\end{eqnarray}
With the help of this full holon Green's function $g({\bf k},\omega)$
and spinon Green's function $D^{(0)}({\bf k},\omega)$ in Eq. (6), we
obtain the holon and spinon spectral functions $A_{h}({\bf k},\omega)$
and $A_{s}({\bf k},\omega)$. Substituting these spectral functions
into Eq. (5), we therefore can obtain the electron spectral function.

In the $t$-$J$ model, the doubly occupied Hilbert space has been pushed
to infinity as Hubbard $U\rightarrow \infty$ and therefore the spectrum
function only describes the lower Hubbard band. Although the particular
details of the electron spectrum and dispersion may differ from compound
to compound, some qualitative features, such as the flat band near
momentum ($\pi$,0) point, seem common and have been universally observed
in hole-doped copper oxide materials \cite{n3}. We have performed a
numerical calculation for the electron spectral function at ($\pi$,0)
point, and the results at (a) the doping $\delta =0.06$ and (b)
$\delta =0.10$ for the parameter $t/J=2.5$ in the zero temperature are
plotted in Fig. 1 (solid line). For comparison, the corresponding
self-consistent mean-field results \cite{n26} (dashed line) are also
plotted in Fig. 1. These results indicate that at the mean-field level,
the electron spectrum at ($\pi$,0) point consists of two main parts,
which comes from noninteracting particles. After including the
fluctuation, the mean-field part is renormalized and the spectral
weight has been spread to lower energies, in particular, the sharp
mean-field peak at ($\pi$,0) point near the chemical potential $\mu$
has been split into two peaks, where the renormalization is strongest.
Moreover, the low energy peaks are well defined at all momenta, and the
positions of the dominant peaks in $A({\bf k},\omega)$ as a function of
momentum in (a) the doping $\delta=0.06$ and (b) $\delta=0.10$ for the
parameter $t/J=2.5$ are shown in Fig. 2 (solid line). In comparison
with corresponding mean-field results \cite{n26} (dashed line) in
Fig. 2, it is shown that in accordance with the property of the
electron spectral functions in Fig. 1, the mean-field electron
dispersion $E^{(0)}_{k}$ in the vicinity of ($\pi$,0) point has been
split into two branches $E^{(-)}_{k}$ and $E^{(+)}_{k}$, and a
pseudogap opens. The branch $E^{(-)}_{k}$ has a very weak dispersion
around ($\pi$,0) point, and then the flat regime appears, while the
Fermi energy is only slightly above this flat regime. This result is
in agreement with those obtained within the concept of proximity of
the underdoped regime to electronic topological transition \cite{n17}.
The anomalous electron dispersion in the present theoretical framework
is determined by the strong electron correlation which give rise to the
local holon-spinon correlation, and is also consistent with the
numerical simulations in low temperatures \cite{n15,n16}. The momentum
dependence of the pseudogap is in qualitative agreement with the
experiments \cite{n5,n6,n7,n8} in that its value, which is of the
order of the magnetic exchange energy $J$, occurs around ($\pi$,0)
point. On the other hand, our results also indicate that although the
electron spectrum is changed with dopings in the underdoped regime,
the flat band near ($\pi$,0) point and pseudogap still exist in a wide
range of dopings. We emphasize that the simplest $t$-$J$ model can
not be regarded as the complete model for the quantitative comparison
with copper oxide materials, but our present results only are in
qualitative agreement with the experiments.

Although the nature of the pseudogap is different in different
theories, the present results show that the pseudogap near ($\pi$,0)
point is closely related to the spinon fluctuation, since the full
holon Green's function (then the holon and electron spectral
function) is obtained by considering the second-order correction due
to the spinon pair bubble, where the single-particle hopping is
strongly renormalized by the short-range AF order resulting in a
bandwidth also of order of (a few) $J$, this renormalization is then
responsible for the anomalous dispersion around ($\pi$,0) point and
normal-state pseudogap. It has been shown that an remarkable point of
the pseudogap is that it appears in both of spinon and holon
excitations. We \cite{n24,n25} have found that this saddle-point
singularity at ($\pi$,0) point and normal-state pseudogap also lead to
the holon pseudogap, which is responsible for the metallic to
semiconducting crossover in the c-axis resistivity $\rho_{c}$ and the
deviation from the temperature linear behavior in the in-plane
resistivity $\rho_{ab}$ in the underdoped regime in copper oxide
materials. In other words, the unconventional normal-state transport
properties in the underdoped regime are attributed to the presence
of the saddle-point singularity around ($\pi$,0) point and the
normal-state pseudogap.

In summary, we have discussed the anomalous momentum and doping
dependence of the electron spectrum and electron dispersion of copper
oxide materials in the underdoped regime within the $t$-$J$ model.
It is shown that the electron spectrum is changed with dopings, and
the electron dispersion exhibits the flat band around ($\pi$,0) point,
which leads to the pseudogap formation. Our theoretical results are
consistent with the experiments and numerical simulations.

Finally, we note that the angle-resolved photoemission spectroscopy
has been carried out \cite{n29} on $(La_{1.28}Nd_{0.6}Sr_{0.12})
CuO_{4}$, a model system of the charge- and spin-ordered state, or
stripe phase, where the electron dispersion also exhibits a flat band
around ($\pi$,0) point. This behavior is consistent with $(La_{2-x}
Sr_{x})CuO_{4}$ near the metal-insulator transition region \cite{n29}
of $x\sim 0.05$ to 0.07. In this doped regime, the holon kinetic
energy is much smaller than the spinon magnetic energy, {\it i.e.},
$\delta t\ll J$, then it is possible that the flat band behavior is
dominated by the strong magnetic fluctuation with AFLRO. This and
other related issues are under investigation now.

\vskip 2cm
\centerline{Acknowledgements}

The authors would like to thank Professor C.D. Gong, Professor H.Q.
Lin, and Professor Z.X. Zhao for helpful discussions. This work was
supported by the National Natural Science Foundation under Grant No.
19774014 and the State Education Department of China through the
Foundation of Doctoral Training.

\newpage
\centerline{Figures}

FIG. 1. Spectral function $A({\bf k},\omega )$ at ($\pi$,0) point for
(a) the doping $\delta =0.06$ and (b) $\delta =0.10$ for the parameter
$t/J=2.5$ in the zero temperature. The dashed line is the result at the
mean-field level.

FIG. 2. Position of the the dominant peaks in $A({\bf k},\omega )$ as
a function of momentum at (a) the doping $\delta=0.06$ and (b)
$\delta=0.10$ for the parameter $t/J=2.5$. The dashed line is the
result at the mean-field level.


\newpage
\begin{enumerate}

\bibitem [*] {add} Mailing address.

\bibitem {n1} A. P. Kampf, Phys. Rep. {\bf 249}, 219 (1994), and
references therein.

\bibitem {n2} M.A. Kastner {\it et al}., Rev. Mod. Phys. {\bf 70},
897 (1998), and references therein.

\bibitem {n3} Z.X. Shen and D.S. Dessau, Phys. Rep. {\bf 253}, 1
(1995), and references therein.

\bibitem {n4} Z.X. Shen {\it et al}., Sciences {\bf 267}, 343 (1995).

\bibitem {n5} D.S. Dessau {\it et al}., Phys. Rev. Lett. {\bf 71},
2781 (1993).

\bibitem {n6} D.S. Marshall {\it et al}., Phys. Rev. Lett. {\bf 76},
4841 (1996).

\bibitem {n7} B.O. Wells {\it et al}., Phys. Rev. Lett. {\bf 74},
964 (1995).

\bibitem {n8} K. Gofron {\it et al}., Phys. Rev. Lett. {\bf 73}, 3302
(1994).

\bibitem {n9} P.J. White {\it et al}., Phys. Rev. B {\bf 54}, R15669
(1996); A.G. Loeser {\it et al}., Science {\bf 273}, 325 (1996).

\bibitem {n10} H. Ding {\it et al}., Nature {\bf 382}, 51 (1996); H.
Ding {\it et al}., Phys. Rev. B {\bf 54}, R9678 (1996).

\bibitem {n11} T. Sato {\it et al}., Phys. Rev. Lett. {\bf 83}, 2254
(1999).

\bibitem {n12} Z.X. Shen and J.R. Schrieffer, Phys. Rev. Lett.
{\bf 78}, 1771 (1997).

\bibitem {n13} A.V. Balatsky and Z.X. Shen, Science {\bf 284}, 1137
(1999).

\bibitem {n14} N. Bulut, D.J. Scalapino, and S.R. White, Phys. Rev.
B {\bf 50}, 7215 (1994).

\bibitem {n15} P. Preuss {\it et al}., Phys. Rev. Lett. {\bf 79},
1122 (1997).

\bibitem {n16} E. Dagotto and A. Nazarenko, Phys. Rev. Lett. {\bf 73},
728 (1994).

\bibitem {n17} F. Qunfrieva and P. Pfeuty, Phys. Rev. Lett. {\bf 82},
3136 (1999).

\bibitem {n18} W.G. Yin, C.D. Gong, and P.W. Leung, Phys. Rev. Lett.
{\bf 81}, 2534 (1998).

\bibitem {n19} P.W. Anderson, J. Phys. Chem. Solids {\bf 54}, 1073
(1993); R.B. Laughlin, Phys. Rev. Lett. {\bf 79}, 1726 (1997).

\bibitem {n20} Shiping Feng, Z.B. Su, and L. Yu, Phys. Rev. B
{\bf 49}, 2368 (1994); Mod. Phys. Lett. B{\bf 7}, 1013 (1993).

\bibitem {n21} P.W. Anderson, in "{\it Frontiers and Borderlines
in Many particle Physics}", edited by R.A. Broglia and J.R.
Schrieffer (North-Holland, Amsterdam, 1987)p. 1; Science {\bf 235},
1196 (1987); F.C. Zhang and T.M. Rice, Phys. Rev. B {\bf 37},
3759 (1988).

\bibitem {n22} E. Dagotto, Rev. Mod. Phys. {\bf 66}, 763 (1994);
T.M. Rice, Physica C {\bf 282-287}, xix (1997).

\bibitem {n23} L. Zhang, J.K. Jain, and V.J. Emery, Phys. Rev. B
{\bf 47}, 3368 (1993); Shiping Feng {\it et al}., Phys. Rev. B
{\bf 47}, 15192 (1993).

\bibitem {n24} Shiping Feng and Zhongbing Huang, Phys. Lett. A
{\bf 232}, 293 (1997); Zhongbing Huang and Shiping Feng, Mod. Phys.
Lett. B {\bf 12}, 735 (1998).

\bibitem {n25} Shiping Feng {\it et al}., Phys. Rev. B {\bf 60},
7565 (1999); and unpublished.

\bibitem {n26} Shiping Feng and Yun Song, Phys. Rev. B {\bf 55}, 642
(1997).

\bibitem {n27} D.M. King {\it et al}., Phys. Rev. Lett. {\bf 73},
3298 (1994).

\bibitem {n28} D.M. Newns {\it et al}., Comments Condens. Matter
Phys. {\bf 15}, 273 (1992), and references therein.

\bibitem {n29} X.J. Zhou {\it et al}., Science {\bf 286}, 268 (1999).

\end{enumerate}
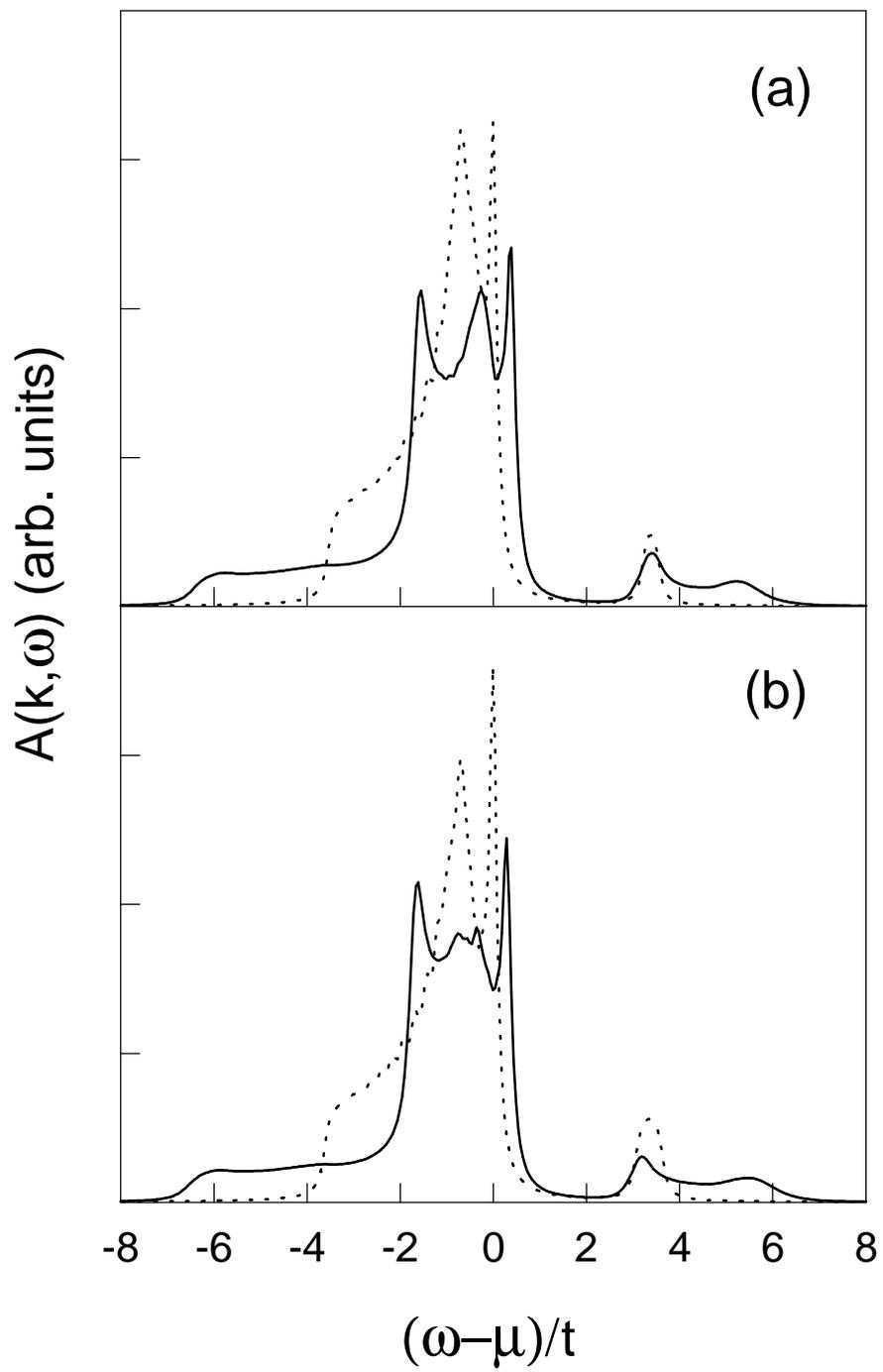
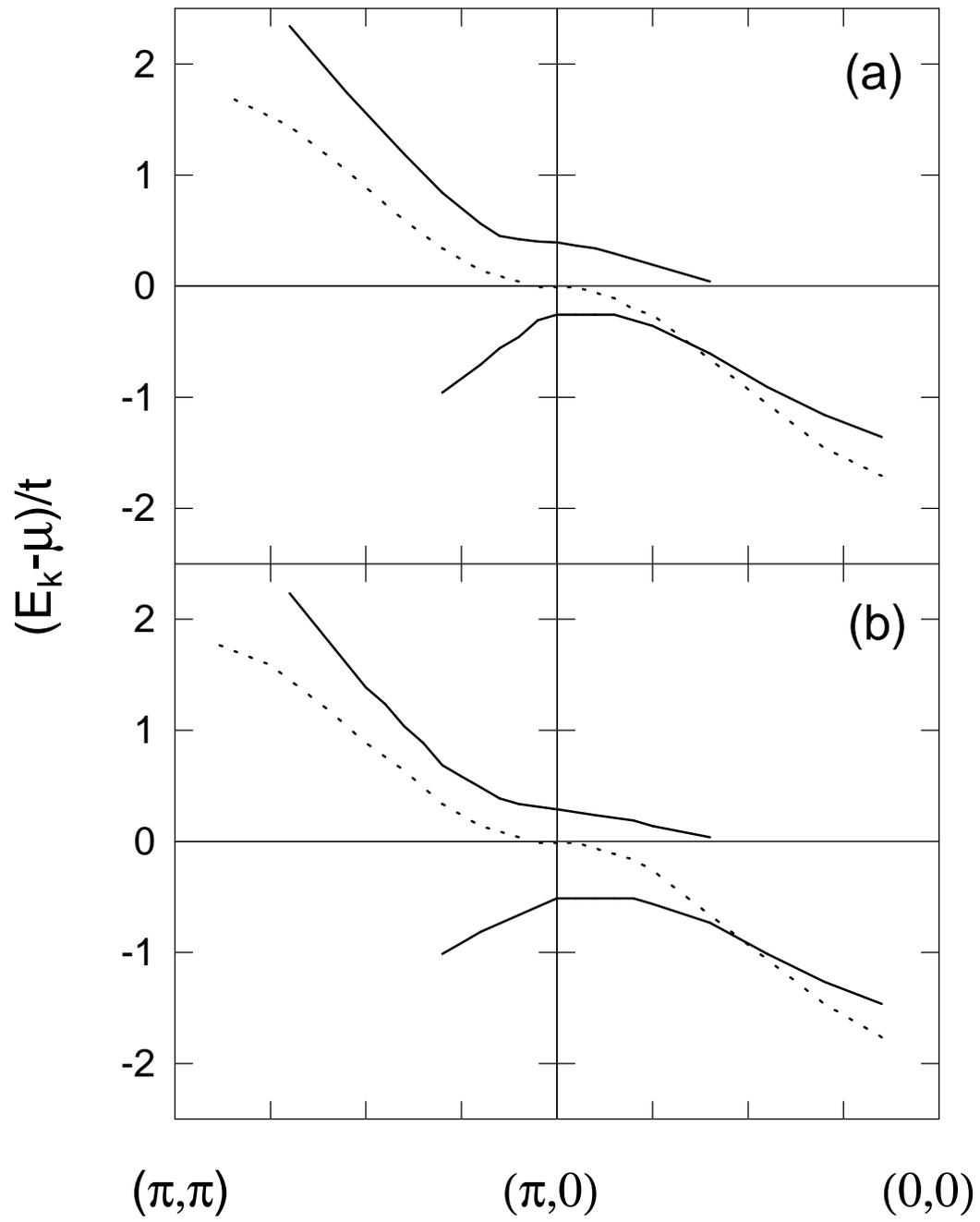
\end{document}